\documentclass[5p,twocolumn]{elsarticle}

\usepackage{lineno,hyperref,amsmath,amssymb,mathrsfs,graphicx}
\usepackage{color}

\journal{Physica D}

\begin{document}

\begin{frontmatter}

\title{Time-dependent spectral renormalization method}
\author{Justin T. Cole and Ziad H. Musslimani}
\address{Department of Mathematics, Florida State University, Tallahassee, FL 32306-4510}
\date{\today}
\begin{abstract}
The spectral renormalization method was introduced by Ablowitz and Musslimani in 2005, 
[Opt. Lett. {\bf 30}, pp. 2140-2142] as an effective way to numerically compute (time-independent) bound states for certain nonlinear boundary value problems.
In this paper, we extend those ideas to the {\it time} domain and introduce a {\it time-dependent spectral renormalization} method as a numerical means to simulate linear and nonlinear evolution equations. The essence of the method is to convert the underlying evolution equation from its partial or ordinary differential form (using Duhamel's principle) into an integral equation. The solution sought is then viewed as a fixed point in both space and time. The resulting integral equation is then numerically solved using a simple renormalized fixed-point iteration method. Convergence is achieved by introducing a {\it time-dependent} renormalization factor which is numerically computed from the {\it physical properties} of the governing evolution equation. The proposed method has the ability to incorporate physics into the simulations in the form of conservation laws or dissipation rates. This novel scheme is implemented on benchmark evolution equations:  the classical nonlinear Schr\"odinger (NLS), integrable $PT$ symmetric nonlocal NLS and the viscous Burgers' equations, each of which being a prototypical example of a conservative and dissipative dynamical system. Numerical implementation and algorithm performance are also discussed. 
 \end{abstract}
\end{frontmatter}
\section{Introduction}
 Computational methods play an indispensable role in various branches of the physical \cite{comp_phys_book}, chemical and biological \cite{comp_neuro_book, comp_bio_book} sciences. In many cases the underlying phenomenon being studied is modeled by either a single or a set of ordinary or partial differential equations. Well-known examples include the quantum Schr\"odinger equation governing the time evolution of a quantum wave function \cite{quantum_book}, reaction-diffusion type systems \cite{math_bio_book} that describe chemical reaction or population dynamics (such as swimming microorganism), and the Navier-Stokes equation modeling the motion of an incompressible Newtonian fluid \cite{fluid_dynam_book1} to mention a few. 

Over the years, numerical tools have played an ever increasing role in the advancement of scientific discoveries. They provide a unique opportunity to tackle many challenging scientific problems that are otherwise difficult to solve. This is the case, for instance,  in complex turbulent flows, weather prediction, stochastic neural dynamics and many body physics with large degrees of freedom. Surprisingly enough, computational methods and numerical simulations can actually ignite new fundamental ideas and ultimately lead to the development of new theories. One example that stands out is the concept of a soliton which emerged as a result of numerical experiments. Solitons, shape-invariant nonlinear waves that exhibit particle-like behavior upon collision, were discovered in 1965 by Zabusky and Kruskal \cite{Zabusky} while performing numerical simulations on the Korteweg-de Vries equation \cite{Korteweg}. Their findings sparked intense research interest in many areas of physics and mathematics which subsequently led to the establishment of the inverse scattering transform and integrable nonlinear evolution equations \cite{Ablowitz_book_III,AKNS}. Soon thereafter, the notion of soliton or solitary wave, spread to many diverse areas in the physical, chemical and biological sciences. Examples include, optical spatial and temporal solitons \cite{Kivshar_book, Malomed}, atomic Bose-Einstein condensates \cite{Pethick_book, Perez_BEC}, atomic chains \cite{Sievers}, molecular and biophysical systems \cite{Davydov} and electrical lattices \cite{Marquii}.

A unifying theme among such diverse fields is the development of computational methods capable of accurately and efficiently capturing the physics under study. As such, numerous numerical tools have been developed in the last few decades to simulate evolution equations of the nonlinear Schr\"odinger and Gross-Pitaevskii type 
\cite{Antoine,Bao,Chang,Muruganandam,Vudragovic,Javidi,Wang}. Perhaps the simplest schemes are finite difference methods such as the Crank-Nicolson algorithm \cite{Chang, Muruganandam, Vudragovic}. Among the easiest to implement are the well-known Runge-Kutta integrators \cite{Javidi}. For evolution equations that can be written as a coupled system of linear and nonlinear equations, both of which can be solved exactly, time-splitting methods are fast and typically quite accurate \cite{Antoine, Wang}. For stiff problems the so-called exponential differencing methods \cite{Hochbruck,Fornberg} are effective. Here, the underlying evolution equation is written in an integral form which is then solved using exponential time-stepping schemes \cite{Beylkin, cox_matthews, trefethen}. 
An issue of great importance is how to devise numerical schemes capable of enforcing physics into the simulations. For example, when it comes to simulating a conservative dynamical system one faces the challenge of imposing conservation laws. In the time-independent setting multiple methods are known that yield stationary soliton solutions that conserve total power 
\cite{Bao1, Perez_Sobolev}. If the system under study happens to be Hamiltonian, a commonly used method is the geometric or symplectic integrators \cite{geometric, Schober, Perez_Symplectic}. They are capable of exactly preserving the symplectic area in phase space and the Hamiltonian. 

In this paper we introduce the {\it time-dependent spectral renormalization} method as an effective and simple computational tool to numerically simulate linear and nonlinear evolution equations. The proposed algorithm has the capability to incorporate the underlying laws of physics in the form of conservation laws or dissipation/rate equations. The idea is to convert the given dynamical system from its evolution equation into an integral form. This approach allows us to think of the solution sought as being a fixed point in space and time of the resulting integral equation. To compute the fixed point a {\it time-dependent} renormalization factor is introduced which is found using the {\it physical properties} of the original governing evolution equation such as conservation laws or dissipation rates. The solution is then obtained from a renormalized fixed-point iteration scheme. This novel time-dependent spectral renormalization approach allows the flexibility to ``integrate" physics into the simulations. The proposed method is applied on benchmark problems: the classical nonlinear Schr\"odinger, integrable $PT$ symmetric nonlocal nonlinear Schr\"odinger and viscous Burgers' equations, each being a prototypical representative of conservative and dissipative evolution equations. Our scheme extendes the {\it time-independent} spectral renormalization concept introduced by Ablowitz and Musslimani in 2005 \cite{spec_renorm_AM} to the {\it time} domain. We point out that the steady state spectral renormalization has been widely used in many applications related to nonlinear optics \cite{yang_vortex, yang_vortex2, Ilan, Ilan2}, Bose-Einstein condensation \cite{Hoefer} and water waves \cite{AFM}. 

The paper is organized as follows. In Sec.~\ref{time-dep-spec} we introduce the time-dependent spectral renormalization scheme and detail its implementation to general evolution equations. The numerical implementation of the algorithm to conservative systems is given in 
Sec.~\ref{conserve_sec} where the classical and integrable $PT$ symmetric NLS equations are used to study important properties of the method. Application of the scheme to dissipative evolution equations is presented in Sec.~\ref{dissipate_sec} where the Burgers' equation is used as a benchmark problem to highlight the generalities of the proposed scheme. We conclude in Sec.~\ref{conclude}.
\section{Time-dependent spectral renormalization}
\label{time-dep-spec}
To begin, we consider a complex or real valued wave function $\psi (x,t)$ that depends on 
space variable $x$ and time $t\ge 0.$ The time evolution of $\psi$ is given by 
\begin{equation}
\label{time-field}
\frac{\partial\psi}{\partial t} = \mathcal{L} \psi  + \mathcal{N} [\psi ] \;, 
\end{equation}
where $\mathcal{L}$ is a linear differential operator and $\mathcal{N}$ is a nonlinear function of 
$\psi.$ The spatial domain $\Omega$ over which the partial differential 
equation (\ref{time-field}) is defined can be either bounded or unbounded. Like most physically relevant evolution equations, Eq.~(\ref{time-field}) represents either a conservative or dissipative dynamical system. In the former case, this implies the existence of conserved quantities given by
 \begin{equation}
\label{conserve-quant}
\int_\Omega Q_j[\psi (x,t)] dx = C_j \;,
\end{equation}
where all $C_j,  j=1,2,3,\cdots$ are constant in time. In the latter case, a set of dissipation or rate equations induced from the dynamical system (\ref{time-field}) are derived in the form
 \begin{equation}
\label{diss-quant}
\frac{d}{dt}\int_\Omega X_j[\psi (x,t)] dx = \int_\Omega Y_j[\psi (x,t)] dx\;,
\end{equation}
where $X_j$ and $Y_j,  j=1,2,3,\cdots$ are functions of $\psi (x,t)$ and referred to as the density and flux, respectively.
To highlight the effectiveness of our method and emphasize its usefulness, we shall limit the discussion to linear differential operators $ \mathcal{L}$ with constant coefficients given by
\begin{equation}
\label{L-phys}
\mathcal{L}  = \sum_{n=1}^Na_n \frac{\partial^n}{\partial x^n}\;,
\end{equation}
where all $a_n$ are constants in space and time. The treatment of a more general case for which all or some of the coefficients $a_n$ are allowed to depend on $x$ and/or $t$ is also possible. In this situation, any such term(s) would be included in the nonlinearity $\mathcal{N}[\psi ].$ Equation (\ref{time-field}) is supplemented with the initial condition 
\begin{equation}
\label{IC}
\psi (x,t=0) = f(x) \;, 
\end{equation}
and either periodic, $\psi (x+L,t) = \psi (x,t),$ with $L$ being the domain size, or rapidly decaying to zero boundary conditions.
Throughout the rest of the paper, we shall use the symmetric forward and inverse Fourier transforms defined by
\begin{equation}
\label{Fourier-forward-gen}
\hat{\psi}(k,t) = F[\psi (x,t)]  \equiv \frac{1}{\sqrt{2\pi}} \int_{-\infty}^{+\infty}  
\psi (x,t) e^{-ikx}  dx \;,
\end{equation}
\begin{equation}
\label{Fourier-inv-gen}
\psi (x,t) = F^{-1}[\hat{\psi} (k,t)]  \equiv  \frac{1}{\sqrt{2\pi}} \int_{-\infty}^{+\infty} 
\hat{\psi}(k,t) e^{ikx} dk\;.
\end{equation}
For problems formulated on a bounded spatial interval $\Omega = [0,L]$ we instead represent the field $\psi (x,t)$ in terms of its Fourier series representation. 
The implementation of the time-dependent spectral renormalization algorithm to a general evolution equation of the type given in (\ref{time-field}) follows several simple steps which we next outline in detail.
\begin{enumerate}
\item Rewrite Eq.~(\ref{time-field}) in Duhamel's integral form:
\begin{equation}
\label{psi-int-gen}
\psi (x,t) = S(t) f(x) +
\int_{0}^td\tau S(t-\tau) \mathcal{N}\left[ \psi (x, \tau ) \right]\;,
\end{equation}
with $S(t)$ being the so-called time propagator (or semi-group) defined by
\begin{equation}
\label{propagator-def-general}
S(t) \equiv e^{t\mathcal{L}} \;, \;\;\;\;\; t\ge 0 \;.
\end{equation}
Equation (\ref{psi-int-gen}) is often used as a departing point for the derivation of many 
time-stepping schemes such as Runge-Kutta \cite{trefethen_book, fornberg_book, yang_book, Taha} and exponential differencing methods \cite{cox_matthews}, among others. Importantly, formula (\ref{psi-int-gen}) implies that the solution $\psi (x,t)$ can be viewed as a fixed point in space and time. The propagator $S(t)$ can be computed with the aid of the Fourier transform or Fourier series. Thus, we have the following representation
\begin{equation}
\label{propagator-comp-gen}
S(t) \eta (x) = F^{-1}\left[ e^{t\hat{\mathcal{L}}(k)} \hat{\eta}(k) \right] \;,
\end{equation}
where $\eta (x)$ is an arbitrary square-integrable or periodic function and $\hat{\mathcal{L}}(k)$ is the Fourier symbol corresponding to the linear operator $\mathcal{L}$ defined by (here $i=\sqrt{-1}$) 
\begin{equation}
\label{L-four}
\hat{\mathcal{L}}(k)  \equiv \sum_{n=1}^Na_n(ik)^n\;.
\end{equation}
\item Introduce a {\it time-dependent} renormalization factor $R(t)$ by rewriting the function as
\begin{equation}
\label{time-renorm-gen}
\psi (x,t) = R(t) \phi (x,t) \;.
\end{equation}
Here, $ \phi (x,t)$ is either a real or complex valued function depending on the nature of the partial differential equation (\ref{time-field}). At this stage, $R(t)$ is an unknown function of time and is {\it assumed} to be real and nonzero at all times. A remark about complex renormalization is discussed in Sec.~\ref{conserve_sec} in connection to the classical and $PT$ 
symmetric NLS equations. Substituting (\ref{time-renorm-gen}) into Eq.~(\ref{psi-int-gen}) 
gives an integral representation for the new field
\begin{eqnarray}
\label{phi-int-gen}
\phi (x,t) &=& \frac{S(t) f(x)}{R(t)}  
\\ \nonumber
&+& \frac{1}{R(t)} \int_{0}^td\tau S(t-\tau) \mathcal{N}\left[ R(\tau) \phi (x, \tau ) \right]   \;.
\end{eqnarray}
Equation (\ref{phi-int-gen}) constitutes the basis for the spectral renormalization method. The solution $\phi (x,t)$ is numerically found from the fixed-point iteration
\begin{eqnarray}
\label{phi-int-gen-fp}
\!\!\!\!\!\!\!\!\!\!\!\!\!
\phi_{n+1} (x,t)  & \!\! = \!\! & \frac{S(t) f(x)}{R_n(t)}  
 \\ \nonumber
&+&
\!\!\!\!\! \frac{1}{R_n(t)} \int_{0}^td\tau S(t-\tau) \mathcal{N}\left[ R_n(\tau) \phi_n (x, \tau ) \right]\;,
\end{eqnarray}
for  $n=1,2,3, \cdots$. This iteration is seeded with an initial guess $\phi_1(x,t).$ 
\item Compute the renormalization factor $R(t)$ from the associated conservation law(s) 
or dissipation (rate) equation(s) {\it induced} from dynamical system (\ref{time-field}). Specifically speaking, we have the following:
\begin{itemize}
\item
If Eq.~(\ref{time-field}) is conservative, then the renormalization factor is computed 
from (\ref{conserve-quant}), i.e.,
 \begin{equation}
\label{conserve-quant-R}
\int_\Omega Q_j[R(t)\phi (x,t)] dx = C_j \;.
\end{equation}
\item 
If Eq.~(\ref{time-field}) is dissipative, then the renormalization factor is derived from the rate equation (\ref{diss-quant}) which gives an evolution equation for $R(t)$ in the form
 \begin{equation}
\label{diss-quant-R}
\frac{d}{dt}\int_\Omega X_j[R(t)\phi (x,t)] dx 
= \int_\Omega Y_j[R(t)\phi (x,t)] dx\;.
\end{equation}
\end{itemize}
The calculation of $R(t)$ is exemplified in Secs.~\ref{conserve_sec} and \ref{dissipate_sec} when applied to the NLS and viscous Burgers' equations. This in turn allows the simulation ``to keep in touch" with the original evolution equation and incorporate relevant physics into the integrator. 
\item Evaluate the time integral that appears in Eq.~(\ref{phi-int-gen}). For ease of presentation we define the quantity
\begin{equation}
\label{G-gen}
G(x,\tau ) \equiv  \mathcal{N} [ R(\tau) \phi (x,\tau)]\;.
\end{equation}
Thus, the integral we are interested in computing is given by 
\begin{equation}
\label{I-space-gen}
I(x,t) \equiv \int_0^td\tau S(t -\tau ) G(x,\tau ) \;,
\end{equation}
which, in Fourier space, takes the form 
\begin{equation}
\label{I-Fourier-gen}
\hat{I}(k,t) = \int_0^td\tau e^{(t -\tau )\hat{\mathcal{L}}(k)} \hat{G}(k,\tau ) \;.
\end{equation}
To numerically evaluate the integral (\ref{I-Fourier-gen}) we divide the time interval $[0, T]$ with $T$ being the end evolution time into $M$ (even) equally spaced intervals each 
of size $\Delta t \equiv T/M.$ Then, denote the time levels by $t_j \equiv j\Delta t, j = 0,1, 2, 3, \cdots, M.$ Clearly, $t_0 = 0$ (initial time) and $t_{M}=T$ (end evolution time). 
After some algebra, we find that $\hat{I}(k,t_m)$ satisfies the recursion relation
\begin{eqnarray}
\label{I-Fourier-recursion-gen}
 \hat{I}(k,t_{m+1}) &= &e^{\Delta t \hat{\mathcal{L}}(k)}  \bigg[ \hat{I}(k,t_{m}) 
 \\ \nonumber
 & + & \int_{t_m}^{t_{m+1}} 
d\tau e^{(t_m -\tau ) \hat{\mathcal{L}}(k)} \hat{G}(k,\tau ) \bigg] \; ,
\end{eqnarray}
valid for $m = 0,1, 2, \cdots , M-1.$ Our goal next is to derive an approximate formula for the integral that appears in Eq.~(\ref{I-Fourier-recursion-gen}). This can be accomplished by replacing the integrand $\hat{G}(k,\tau )$ by either a constant, linear, quadratic or higher order polynomial of the time variable $\tau$. This is referred to as Filon integration \cite{Iserles}. See
also \cite{Hochbruck, cox_matthews} for applications of the Filon method in exponential time differencing. In this paper, we present results only for linear interpolants. On each time interval $t_m \le \tau \le t_{m+1}, m = 0,1, 2, \cdots , M-1,$ we interpolate the 
function $\hat{G}(k,\tau )$ by the linear function
\begin{eqnarray}
\label{G-interp-gen}
 \hat{G}(k,\tau ) &=& \hat{G}(k,t_m ) 
\\ \nonumber
& +&
\frac{1}{\Delta t}
\left[\hat{G}(k,t_{m+1} ) - \hat{G}(k,t_m)\right] (\tau - t_m)\;.
\end{eqnarray}
Substituting Eq.~(\ref{G-interp-gen}) into (\ref{I-Fourier-recursion-gen}) and performing integration by parts, we arrive at the result
\begin{align}
\label{I-Fourier-recursion-2-gen}
 \hat{I}(k,t_{m+1}) = ~& e^{\Delta t \hat{\mathcal{L}}(k)} 
\bigg[ \hat{I}(k,t_{m}) \\ \nonumber
& +  A \hat{G}(k,t_m ) + B \hat{G}(k,t_{m+1} ) \bigg], 
\end{align}
for $m = 0,1, 2, \cdots , M-1$ 
with $\hat{I}(k,t_0)=\hat{I}(k,0)\equiv 0.$ Here, we define 
\begin{equation}
\label{quad_coeffs-gen-A}
A = \frac{e^{-\Delta t  \hat{\mathcal{L}}(k)} +  \Delta t  \hat{\mathcal{L}}(k) - 1}
{\Delta t  \hat{\mathcal{L}}^2(k)} \;,
\end{equation}
\\
\begin{equation}
\label{quad_coeffs-gen-B}
B = \frac{1 - e^{-\Delta t  \hat{\mathcal{L}}(k)} \left( \Delta t  \hat{\mathcal{L}}(k) + 1 \right) }
{\Delta t  \hat{\mathcal{L}}^2(k)}   \; .
\end{equation}
Since the coefficients $A$ and $B$ depend only on $\Delta t$ and the Fourier mode $k$ (but not on the solution itself or the iteration index $m$), they are pre-computed only once. For certain problems, such as $\hat{\mathcal{L}}(k) = -k^2$ at $k = 0$, the operator $\hat{\mathcal{L}}(k)$ vanishes at some value(s) of $k$. In this situation formulas (\ref{quad_coeffs-gen-A}) and
 (\ref{quad_coeffs-gen-B}) are undefined. To remedy this, their values are replaced by their corresponding limits 
$A \rightarrow \Delta t /2$ and $B \rightarrow \Delta t /2$ as $\hat{\mathcal{L}}(k) \rightarrow 0$. This idea has already been implemented in the context of exponential time differencing \cite{cox_matthews}. Another way to avoid dividing by zero is to use the Cauchy integral formula \cite{trefethen}.
\item Implement the time-dependent spectral renormalization fixed-point iteration based 
on Eqs.~(\ref{phi-int-gen}) and (\ref{I-space-gen}). Below we write it both in physical and Fourier spaces respectively:
\begin{equation}
\label{phi-int-iterate-phys-gen}
\phi_{n+1} (x,t) = \frac{1}{R_n(t)}  S(t) f(x)
+
\frac{1}{R_n(t)}  I_n (x,t)\;,
\end{equation}
\begin{equation}
\label{phi-int-iterate-Four-gen}
\hat{\phi}_{n+1} (k,t) = \frac{1}{R_n(t)}  e^{t \hat{\mathcal{L}}(k)} \hat{f}(k)
+
\frac{1}{R_n(t)} \hat{I}_n(k, t)\;,
\end{equation}
where $n = 1,2,3, \cdots.$ At every iteration step $n$ the function $I(x,t)$ or alternatively 
$\hat{I}(k,t)$ is obtained from Eq.~ (\ref{I-Fourier-recursion-2-gen}). 
\end{enumerate}
The iteration schemes (\ref{phi-int-iterate-phys-gen}) or (\ref{phi-int-iterate-Four-gen}) are seeded with an ``arbitrary" initial guess $\phi_1(x,t)$ that satisfies the boundary conditions. Typical examples include: (i) the initial condition (\ref{IC}), (ii) $f(x)h(t)$ with $h(t)$ an ``arbitrary" function of time and (iii) a random function in space and time.
Convergence is achieved when the relative error between successive iterations is less than a prescribed level of error tolerance $\epsilon$, i.e., 
$\max_{x,t}| \phi_{n+1} - \phi_n| \le \epsilon$ as $n\rightarrow \infty$. We remark that the 
time-dependent spectral renormalization approach allows one to choose the size of the total time interval $T$ to be {\it large} enough such that the standard ({\it unrenormalized, or $R(t)= 1$}) fixed point iteration 
\begin{equation}
\label{direct-fp}
\psi_{n+1} (x,t) = S(t) f(x) 
+
\int_{0}^td\tau S(t-\tau) \mathcal{N}\left[ \psi_n (x, \tau ) \right]\;,
\end{equation}
fails to converge. So far, we have presented a general framework where a solution to an evolution equation of the type given in Eq.~(\ref{time-field}) can be numerically obtained by a renormalized time-dependent fixed-point iteration. In the next two sections we implement this scheme on {two important and physically relevant examples that represent conservative and dissipative dynamical systems.
\section{Conservative case} 
\label{conserve_sec}
\subsection{Classical NLS equation}
As a prototypical example representing a conservative dynamical system, we consider the classical nonlinear Schr\"odinger  equation \cite{Ablowitz_book}
\begin{equation}
\label{NLS}
\psi_t= i \psi_{xx} + 2 i |\psi|^2\psi \;, 
\end{equation}
where $\psi = \psi (x,t)$ is a complex-valued function of the real variables  $x$ and time $t \ge 0.$ Subscripts indicate partial derivatives with respect to $x$ and $t.$ Equation (\ref{NLS}) is posed on the whole real line and is supplemented with the following boundary and initial conditions respectively: $\psi (x,t)$ goes to zero sufficiently fast as $|x|\rightarrow\infty$ and $\psi (x,t=0) = f(x)$ which is assumed to be smooth and square-integrable on the whole real line.   There are two important reasons to consider the NLS equation as a test bed for the performance of the time-dependent spectral renormalization scheme. The first is due to its wide applications in optics, condensed matter physics (such as Bose-Einstein condensation, superfluidity and superconductivity) and fluid mechanics (deep water waves). The second reason is tied to the fact that Eq.~ (\ref{NLS}) is an integrable evolution equation and admits an infinite number of conserved quantities \cite{Ablowitz_book_III}. Furthermore, it admits the exact moving soliton solution
\begin{equation}
\label{soliton}
\psi_{\rm ex}(x,t) = {\rm sech}(x +2\xi t) e^{-i[\xi x +(\xi^2 -1) t]} \;,
\end{equation}
where $\xi$ is a real constant used to generate momentum. In the case of more general NLS-type equations several spectrally accurate numerical methods exist to find soliton modes \cite{Chang2,Chang3}. For the NLS equation, we have the following three physically relevant conserved quantities:
\begin{align}
\label{NLS-cons-1}
&\text{power:}\;\;\;\;\;\;\;\;\;\;\;\;\;\;\; J_1\left( \psi \right) \equiv \int_{-\infty}^{\infty} dx | \psi |^2
= C_1  \;, \\
\label{NLS-cons-2}
&\text{momentum:} \;\;\; \;\;\; J_2\left( \psi \right) \equiv \text{Im} \int_{-\infty}^{\infty} dx 
\psi  \psi_x^* = C_2  \;,\\
\label{NLS-cons-3}
&\text{Hamiltonian:}\;\;\;  J_3\left( \psi \right) \equiv\int_{-\infty}^{\infty} dx
\left[| \psi_x|^2 - | \psi |^4 \right] = C_3\;,
\end{align}
where all $C_j$ are constants of motion. Next, we list the necessary steps to implement the time-dependent spectral renormalization method. To this end, we have $\mathcal{L} =  i\frac{\partial^2}{\partial x^2}, \hat{\mathcal{L}}(k) = -ik^2, 
\mathcal{N} = 2i |\psi |^2\psi$, the propagator (or semigroup) defined in formula 
(\ref{propagator-def-general}) given by $S(t)= \exp \left( it\frac{\partial^2}{\partial x^2} \right),$ and the renormalized nonlinear term $G(x,t) =  2i  R^3(t )  | \phi (x, t ) |^2 \phi (x, t ).$ 
Solutions to Eq.~(\ref{NLS}), corresponding to initial condition (\ref{IC}) are then computed from the iterative scheme (\ref{phi-int-iterate-Four-gen}) with the renormalization factor $R(t)$ taken from Table~\ref{Tab2} where
\begin{equation}
|| \varphi ||_p \equiv \left(  \int_{-\infty}^{\infty} |\varphi|^p  dx \right)^{1/p} \; .
\end{equation}
Each choice of $R$ would correspond to a certain conservation law.
\bgroup
\def\arraystretch{2}
\begin{table*}[htbp]
\centering
 \begin{tabular}{ |c |  c   |c |  } \hline 
     & $R^2(t)$ & Conserved Quantity \\  \hline
   1.  & $ \frac{C_1}{||\phi  ||_2^2} $ & $J_1$        \\ \hline
   2. & $ \frac{C_2} { \text{Im} \int dx \phi \phi_x^* }$ & $J_2$  \\ \hline
   3. & $\frac{|| \phi_x ||_2^2 \pm \sqrt{|| \phi_x ||_2^4 - 4C_3  ||\phi ||_4^4} }{{2||\phi ||_4^4}}$ & $J_3$   \\ \hline  
   4.  & $ \frac{ C_1 +  C_2}{ ||\phi  ||_2^2  + \text{Im} \int dx \phi \phi_x^*} $ & $  J_1+  J_2$        \\ \hline 
     & $ \frac{\kappa \pm \sqrt{ \kappa^2 - 4 \alpha_3 || \phi ||_4^4 \mathcal{C}}  }{2 \alpha_3 || \phi ||_4^4}, \mathcal{C} = \alpha_1 C_1 + \alpha_2 C_2 + \alpha_3 C_3, $ & $  \alpha_1 J_1 + \alpha_2 J_2 + \alpha_3 J_3 $     \\ 
  5. &$ \kappa = \alpha_1 || \phi ||_2^2 + \alpha_2 \text{Im} \int dx \phi \phi_x^* + \alpha_3 ||\phi_x ||_2^2$  & $\alpha_3 \not= 0$ \\ \hline  
   6.  & $  \sqrt{\frac{C_1 C_2}{||\phi  ||_2^2  ~ \text{Im} \int dx \phi \phi_x^* }} $ & $J_1J_2$        \\ \hline   
   7.  & $ \frac{C_1 || \phi_x ||_2^2 - C_3 || \phi ||_2^2}{C_1 || \phi ||_4^4} $ & $  J_3/J_1$        \\ \hline
   8.  & $ \frac{C_2 || \phi_x ||_2^2 - C_3  \text{Im} \int dx \phi \phi_x^*}{C_2 || \phi ||_4^4} $ & $  J_3/J_2$        \\ \hline
\end{tabular}
\caption{List of renormalization factors and their corresponding conserved quantity.}
\label{Tab2}
\end{table*}
It is obvious that the ``space of options" to single out one ``favorite" formula for the renormalization factor is rather large. The decision as to which expression to pick is based on 
various factors such as convenience, the physics of the problem and the numerical validity of 
that specific formula.
\begin{figure*}
\centering
\includegraphics[scale=.8]{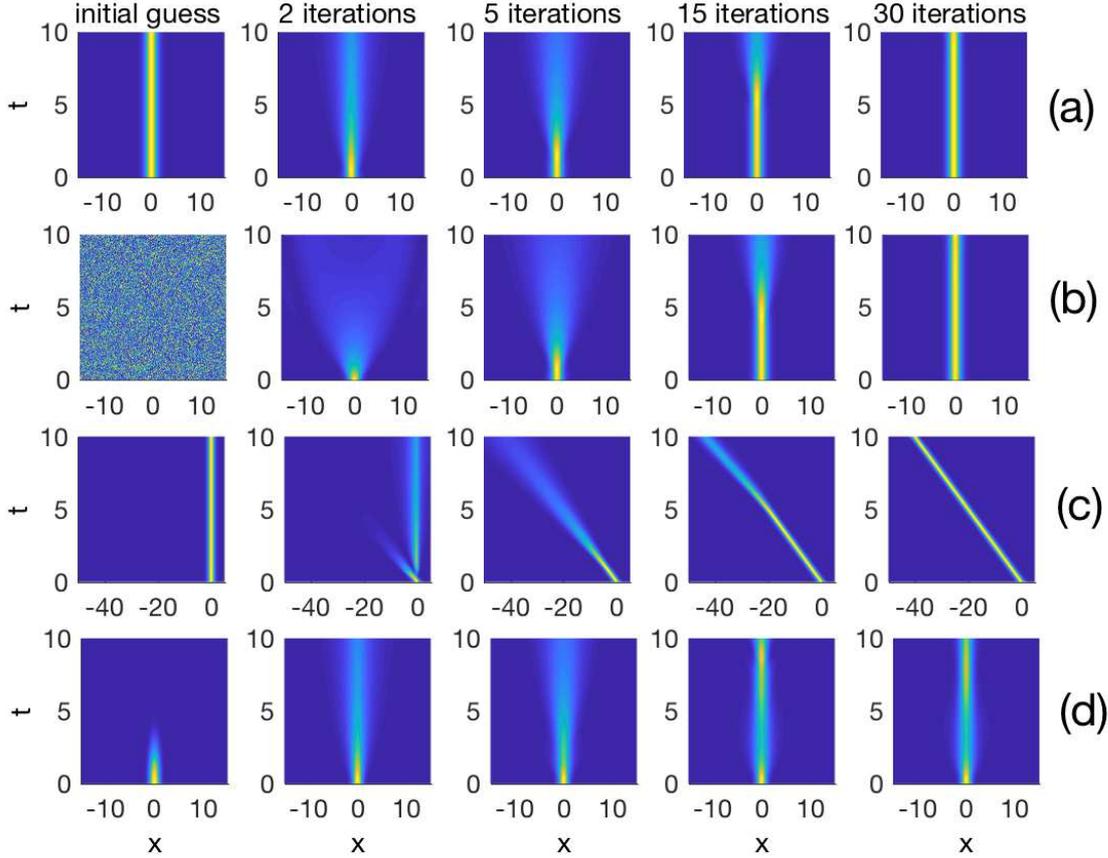}
\caption{A top view snapshots of the intensity $|\psi(x,t)|^2$ at various stages of the iteration process. All results are obtained by iterating 
Eq.~(\ref{phi-int-iterate-Four-gen}) using the renormalization factor given in Table~\ref{Tab2} that exactly conserves the quantity $J_1$, i.e. $R^2(t) = C_1/ || \phi||_2^2$. The NLS initial conditions and iteration initial guesses are: 
(a) $f(x) = {\rm sech}(x), \; \phi_1(x,t) = f(x),$
(b) $f(x) = {\rm sech}(x), \; \phi_1(x,t) = \text{random function of}$ $x$ and $t$ uniformly distributed on the interval $[0,1]\times [0,1].$
(c) $f(x) = {\rm sech}(x) e^{- 2 i x}, \; \phi_1(x,t) = f(x),$
(d) $f(x) = e^{-x^2/2},\; \phi_1(x,t) = e^{-x^2/2} e^{-t^2/10}$. Parameters are: $L=100 \;, N = 1024 \;, M = 200 \;, T= 10.$}
\label{FIG1}
\end{figure*}
\begin{figure} [h]
\includegraphics[scale=.3]{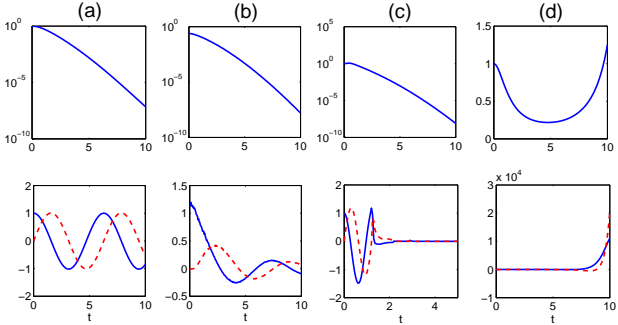}
\caption{Top row shows the dependence of the renormalization factor $R(t) = (C_1/ || \phi||_2^2)^{1/2}$ on time.
Bottom row shows the real (solid blue line) and imaginary (dashed red line) parts of the renormalization factor given in Eq.~(\ref{complex_R_define}). Columns (a), (b), (c), and (d) correspond to the solutions shown in Fig.~\ref{FIG1} rows (a), (b), (c), and (d), respectively.}
\label{FIG2}
\end{figure}
\begin{figure} [h]
\includegraphics[scale=.45]{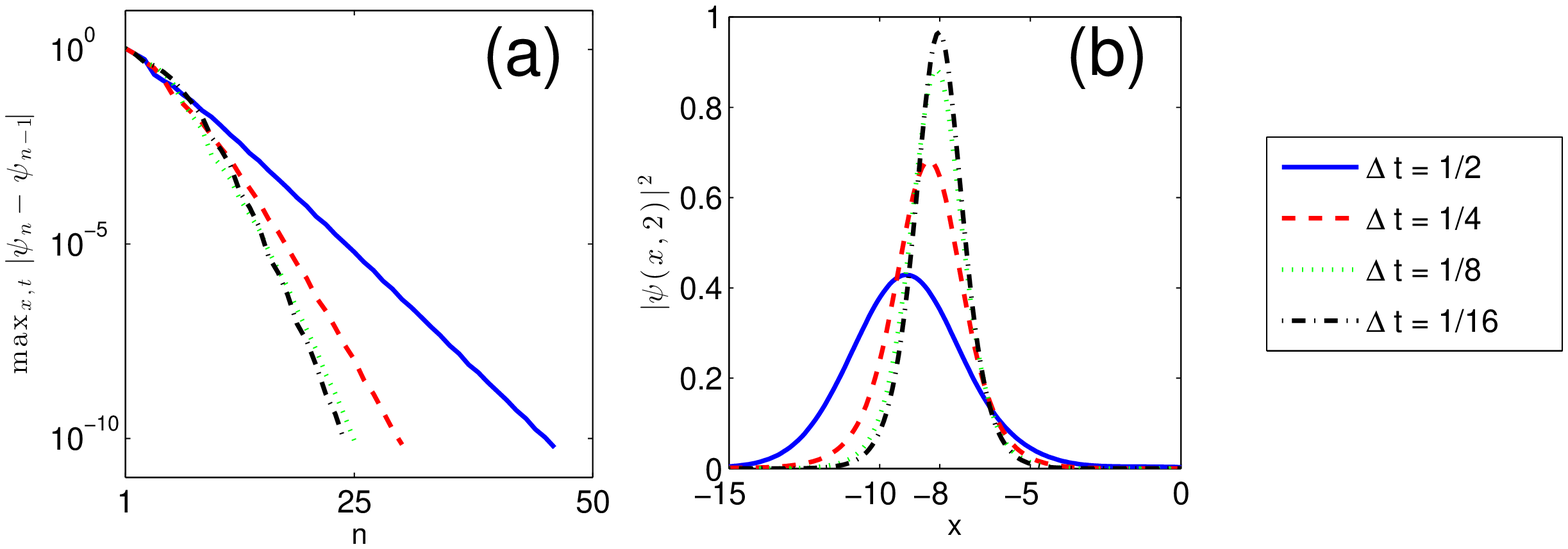}
\caption{(a) Maximum error between successive iterations, 
$\max_{x,t}|\psi_{n+1}(x,t) - \psi_n(x,t)|,$ as a function of number of iterations $n$. (b) 
Intensity $|\psi (x,t=2)|^2$ computed using various values of $\Delta t$. The renormalization factor $R(t)$ is computed using the formula given in row 1 of Table~\ref{Tab2}. Computational parameters are: $\xi = 2, T = 2, L =100, N = 1024$.}
\label{FIG3}
\end{figure}
\begin{figure} [h]
\includegraphics[scale=.45]{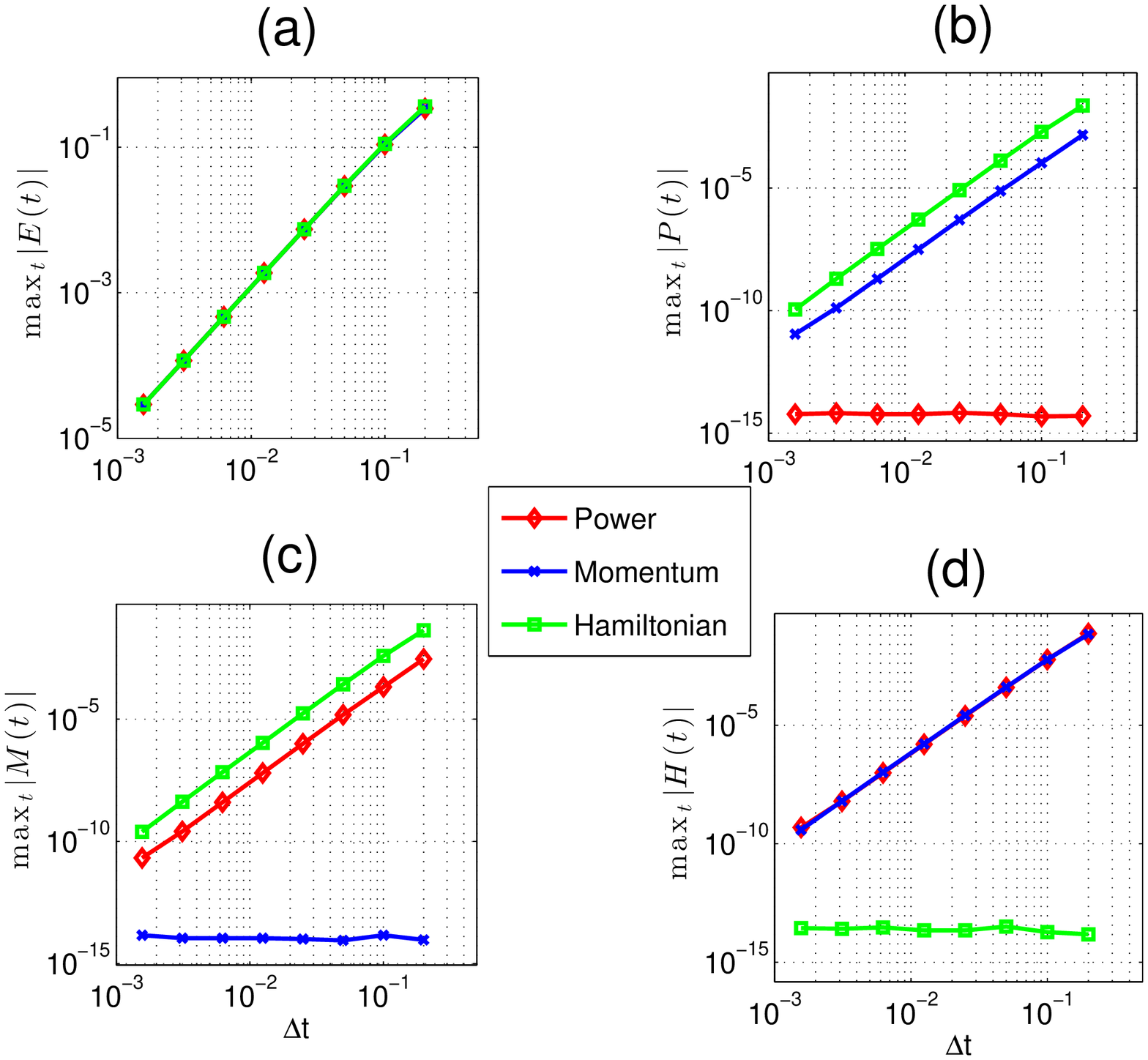}
\caption{Maximum error of the numerically calculated (a) solution, (b) power, (c) momentum, and (d) Hamiltonian. Each curve is obtained using the renormalization factor $R(t)$ by imposing conservation of power (red) taken from row 1 in Table~\ref{Tab2}, momentum (blue) taken from row 2 in Table~\ref{Tab2} and Hamiltonian (green) taken from row 3 ($-$ sign) in 
Table~\ref{Tab2}. The computational parameters are: $\xi = 2, T = 2, L =100, N = 1024$.}
\label{FIG4}
\end{figure}

We have implemented the time-dependent spectral renormalization scheme on the NLS equation corresponding to three different types of initial conditions: 
(i) $f(x) =  {\rm sech}(x)$ which gives rise to a stationary soliton, (ii) 
$f(x) = {\rm sech}(x) e^{- 2 i x}$ leading to a moving pure soliton carrying nonzero 
momentum and (iii) a Gaussian wavepacket $f(x) = e^{-x^2/2}.$ The constants $C_j, j = 1,2,3$ are calculated directly from the initial condition. 

Several iterations of the solution intensity 
$|\psi_n (x,t)|^2$  are shown in Fig.~\ref{FIG1} which is obtained from iterating 
Eqn.~(\ref{phi-int-iterate-Four-gen}) using $R(t)$ that exactly conserves power 
(row 1 in Table~\ref{Tab2}). 
To examine the effectiveness of the method, the iteration is seeded with four different initial guesses. In the first experiment, we chose an initial guess identical to the NLS initial condition, i.e., $\phi_1(x,t) = {\rm sech}(x).$ This choice is somewhat natural to begin with. After a few rounds the wave function has corrected itself and locked on to the soliton solution given in Eq.~(\ref{soliton}) with $\xi =0$ -- see Fig.~\ref{FIG1}(a). To test the scheme's sensitivity to initial guesses, we kept the same initial condition as before but now seeded the iteration with a random field in both space and time chosen to be uniformly distributed on the interval $[0,1]\times [0,1].$  Interestingly enough, the outcome of the run was independent of the initial input as it again converged to the numerical soliton solution. This robust behavior is depicted in Fig.~\ref{FIG1}(b). We next conduct some numerical tests using different initial conditions. Several simulations are shown in Figs.~\ref{FIG1}(c) and (d) for a traveling soliton and Gaussian solutions, respectively. While the results shown in this paper are carried out using conservation of power only (as is the case in Fig.~\ref{FIG1}) we point out that other conservation quantities have been used to converge to the same solution.  All numerical tests have been performed on a large spatial domain $(L=100)$ to insure the problem is indeed posed on the whole real line and guarantee decaying (zero) boundary conditions at all times. The total time interval used to produce Fig.~\ref{FIG1} is $T=10$ which, {\it without} the use of any renormalization technique, would cause the iterative method in Eq.~(\ref{direct-fp}) to {\it diverge}. 

 One interesting behavior we observe is that the renormalization factor approaches zero when dealing with large time intervals. This is the case, for example, when computing stationary solitons $(\xi =0)$ while using real $R(t).$ Figure \ref{FIG2} (a), (b) and (c) depicts such behavior. However, this is not the case when solving the NLS equation with Gaussian input. Here, the (real) renormalization factor remained on the same order as it started for all times -- see  Figure \ref{FIG2} (d). 
 We point out that the spectral renormalization scheme converges even when one chooses a rather large time grid spacing $\Delta t.$  Figure \ref{FIG3} shows maximum error between successive iterations defined by $\max_{x,t} | \psi_{n+1}(x,t) - \psi_{n}(x,t) |$ as a function of the iteration number $n.$ For small $\Delta t$ the scheme converged after some $25$ iterates giving rise to an accurate numerical solution. Interestingly, the iteration converged even for a relatively large $\Delta t$ (on the order of 0.5), however, the resulting solution shows poor accuracy.

To quantify the performance of the scheme, such as convergence, we have measured at each simulation run the maximum spatial difference between the (convergent) numerically obtained solution $\psi_{\rm num}(x,t)$ and the exact soliton given in (\ref{soliton}). Thus, we define 
\begin{equation}
\label{max_norm}
\mathsf{E}(t) = \max_{x} | \psi_{\rm ex}(x,t) - \psi_{\rm num}(x,t) | \; .
\end{equation}
Furthermore, at each simulation run, we have monitored the difference between each numerically computed conserved quantity and its initial value as a function of time. 
Precisely, we checked the time evolution of the following quantities:
\begin{equation}
\label{P}
\mathsf{P}(t) = \int_{-\infty}^{+\infty} dx | \psi_{\rm num}(x,t) |^2 - C_1\;, 
\end{equation}
\begin{equation}
\label{M}
\mathsf{M}(t) = \text{Im} \int_{-\infty}^{+\infty} dx \psi_{\rm num}(x,t)  
\frac{\partial \psi^*_{\rm num}(x,t)}{\partial x}  - C_2\;,
\end{equation}
\begin{equation}
\label{H}
\mathsf{H}(t) = \int_{-\infty}^{+\infty} dx
\left[\left | \frac{\partial \psi_{\rm num}(x,t)}{\partial x}\right |^2 
- | \psi_{\rm num}(x,t) |^4 \right] - C_3\;.
\end{equation}
For a fixed grid size $\Delta t$ we ran three different types of simulations all using the same NLS initial condition $f(x) = {\rm sech}(x) e^{- 2 i x}$ ($\xi =2$), but different $R(t)$ that are computed from either conservation of power, momentum, or Hamiltonian. Once convergence is achieved, the output $\psi_{{\rm num}}(x,t)$ is recorded and used to compute the maximum value of the quantities $\mathsf{E} (t), \mathsf{P}(t), \mathsf{M}(t)$ and $\mathsf{H}(t)$ given in 
Eqns.(\ref{max_norm})--(\ref{H}) over the entire time domain. We then doubled the number of grid points (halved the grid size $\Delta t$) and repeated the same experiment again. This procedure was performed eight times (corresponding to eight different $\Delta t$). 
The findings are summarized in Fig.~\ref{FIG4}.  As one can see from Fig.~\ref{FIG4}(a), the maximum error in the solution is converging to the exact soliton and is doing so at a second-order rate. Furthermore, Fig.~\ref{FIG4}(b), (c) and (d) show the errors in the power, momentum and Hamiltonian to be near machine precision when computed using their respective renormalization factor alone for each $\Delta t$. Otherwise, the error does not seem to be exceptional.
\begin{figure} [h]
\includegraphics[scale=.38]{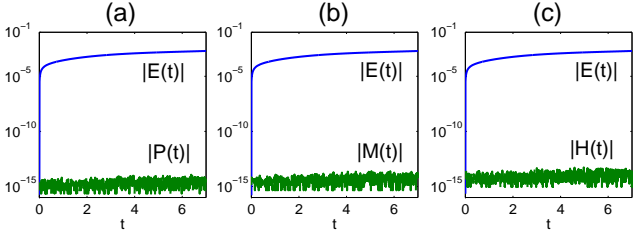}
\caption{The error evolution in the solution $\mathsf{E}(t)$, power $\mathsf{P}(t)$, 
momentum $\mathsf{M}(t)$, and Hamiltonian $\mathsf{H}(t)$. The value of $R(t)$ is found using conservation of power (row 1 of Table~\ref{Tab2}), momentum (row 2 of Table~\ref{Tab2}), and Hamiltonian (row 3 of Table~\ref{Tab2} with $-$ sign) in panels (a), (b), and (c), respectively. The computational parameters are: $\xi = 2, T = 7, M = 1000, L =100, N = 1024$.}
\label{NLS_error_evolve}
\end{figure}
Finally, we have monitored the time evolution of the quantities $\mathsf{E} (t), \mathsf{P}(t), \mathsf{M}(t)$ and $\mathsf{H}(t)$ given in Eqns.~(\ref{max_norm})--(\ref{H}). They are displayed in Fig.~\ref{NLS_error_evolve} for a traveling wave soliton. The error in the computed solution compared to the exact one, $E(t)$, is found to grow with time. 
Importantly, the errors in power, momentum and Hamiltonian are machine precision accurate when computed using their respective renormalization factor (see Fig.~\ref{NLS_error_evolve}). }
\subsection{Comparison to other methods}
\begin{figure} [h]
\includegraphics[scale=.4]{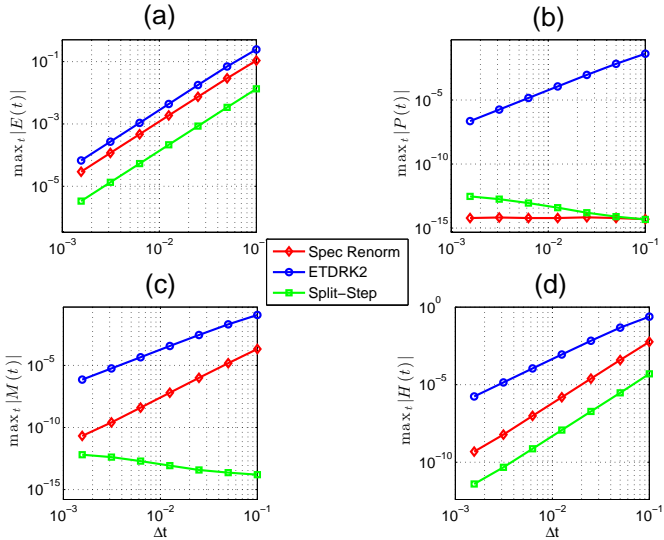}
\caption{Global maximum error of the numerically calculated (a) solution, (b) power, (c) momentum, and (d) Hamiltonian. Curves correspond to numerical solutions obtained through the spectral renormalization, split-step, and exponential time-differencing Runge-Kutta (ETDRK2) methods. Conservation of power is used to determine $R(t)$ -- see row 1 
in Table~\ref{Tab2}. The computational parameters are: $\xi = 2, T = 2, L =100, N = 1024$.}
\label{FIG5}
\end{figure}
It is of considerable interest to compare the time-dependent spectral renormalization method to other well known time-integrators. For that purpose, we choose to solve the NLS equation (\ref{NLS}) with initial condition $f(x) = {\rm sech}(x) e^{- 2 i x}$ using two different second-order accurate schemes: split-step Fourier and exponential time-differencing Runge-Kutta (ETDRK2). In our spectral renormalization simulations we enforced conservation of power only and computed the renormalization factor $R(t)$ from row 1 in Table~\ref{Tab2}. To quantify the performance of each scheme, we repeat the calculations that were performed in Fig. 4 using the three schemes mentioned above. In Fig.~\ref{FIG5}(a) the solution errors obtained from all three integrators are observed to be on the same order of magnitude (with the split-step error slightly lower than the others) and converging at the same rate. The story is different when it comes to conservation of power. Here, both the spectral renormalization and split-step methods are found to be numerically exact (see in Fig.~\ref{FIG5}(b)). Note that the split-step method and certain finite-difference schemes are known to conserve power exactly \cite{Weideman_Herbst,Perez_Conserve}. The error obtained using these two methods are favorable in comparison to exponential time-differencing scheme. For the parameters considered here, the spectral renormalization method yields power values that are around seven orders of magnitude more accurate. On the other hand, the momentum and Hamiltonian are most accurate when computed using the split-step method. We point out that for the ETDRK2 scheme considered here, the physical quantities that any solution must satisfy appear to be constrained by the accuracy of the numerical solution. This is clearly not the case for the spectral renormalization scheme which can enforce certain relevant physical properties to excellent accuracy regardless of how accurate the solution is.

In comparison to standard method of lines type integrators, the spectral renormalization algorithm requires considerable storage for large $t.$ Specifically, one must store $(M+1)N$ data points, where $N, M$ denote the number of spatial and temporal grid points. In two spatial dimensions this storage requirement becomes $(M+1)N^2$. To alleviate this storage burden, as well as to increase the algorithm speed, one possibility is to divide the full time interval $[0,T]$ into $\mathcal{M}$ sub-intervals $[0,T] = [0,T_1] \cup [T_1,T_2] \cdots \cup [T_{M-1},T_M]$. This cuts the required memory inside each sub-interval to $(M+1)N/\mathcal{M}$. 
Each time interval length is chosen sufficiently large such that the spectral renormalization algorithm is effective, efficient and fast. On the time span $[0, T_1]$ the proposed algorithm is implemented using the initial condition $\psi (x,0)=f(x)$ and continues until convergence to a fixed point in space and time is reached. The scheme is then applied to the time interval $[T_1, T_2]$ with initial condition $f(x) = \psi (x,t=T_1).$ The code is repeated until the whole time span is covered. Typically, for problems integrated on time interval $[0,100]$ ten partitions would be sufficient for good performance. Finally, using the fast Fourier transform to approximate derivatives results in CPU run times that are respectable. For example, to reach the converged solutions shown in Fig.~\ref{FIG1} on a standard laptop using Matlab took about one second. Different initial guesses resulted in similar run times.

\subsection{Complex Renormalization Factor}
So far, we have implemented and examined the performance of the time-dependent spectral renormalization scheme under the assumption that $R(t)$ is real valued. In fact this restriction is not needed for the formulation of the scheme. It is rather rooted in the fact that all NLS conservation laws give the magnitude but not the phase of the renormalization factor. To remedy this issue, we propose here one possibility of lifting this constraint by deriving an expression for the renormalization factor directly from the NLS equation. With this approach, one is still able to incorporate conservation laws. To this end, we substitute $\psi (x,t)=R(t)\phi (x,t)$ into Eq.~(\ref{NLS}), integrate over the entire real line while utilizing the fast decay of the wave function (localized boundary conditions) and obtain
\begin{equation}
\label{complex_R_define}
R(t) = \frac{R(0) \alpha(0)}{\alpha(t)} \exp\left[ 2 i  \int_0^t 
\frac{\beta(\tau) |R(\tau )|^2}{\alpha(\tau)} d\tau \right]\;,
\end{equation}
where
\begin{equation}
\label{alpha}
\alpha(t) = \int_{-\infty}^\infty \phi(x,t) dx  \;,
\end{equation}
\begin{equation}
\beta(t) =  \int_{-\infty}^\infty |\phi(x,t)|^2 \phi(x,t)  dx \;.
\end{equation}
The constant $\alpha (0)$ is computed from (\ref{alpha}) whereas $R(0)$ is found from Eq.~(\ref{time-renorm-gen}) and (\ref{IC}). The result is
\begin{equation}
\label{R0}
R(0) =  \int_{-\infty}^\infty \frac{|f(x)|^2}{f^*(x)\phi(x,0)}  dx \;.
\end{equation}
Formula (\ref{complex_R_define}) provides an alternative expression for the renormalization factor $R$ {\it without} any assumption of being real. However, it is valid so long the quantity given in (\ref{alpha}) is not zero. Furthermore, this approach restricts the choices for $\phi (x,0)$ as it could cause $\alpha (0)$ to become zero. Last, but not least, since $\phi$ is in general complex function, the term inside the exponent in Eq.~(\ref{complex_R_define}) can lead to exponential growth or decay which will eventually put serious strain on the time numerical integration. To incorporate physics into (\ref{complex_R_define}) we replace $|R(\tau )|^2$ that appears inside the exponent by any of the expressions given in Table~\ref{Tab2} where $R^2(\tau )$ is now replaced by $|R(\tau )|^2.$ We have implemented the time-dependent spectral renormalization scheme for the complex $R$ case and found that it converged to the solution from a large selection of initial guesses. The main difference between the real and complex renormalization approach is that now $R$ seems not to go to zero; it rather oscillates in time (see Fig.~\ref{FIG2}).
\subsection{$PT$ symmetric integrable nonlocal NLS equation}
In this section we apply the time-dependent spectral renormalization method to the
$PT$ symmetric nonlocal nonlinear Schr\"odinger equation 
\begin{equation}
\label{PTNLS}
\psi_t(x,t) = - i \psi_{xx}(x,t) - 2 i  \psi^2(x,t) \psi^*(-x,t) \; ,
\end{equation}
where $\psi (x,t)$ is a complex valued function of the real variables $x\in\mathbb{R}$ and time
$t\ge 0.$ Equation (\ref{PTNLS}) was first introduced in \cite{Ablowitz1} and shown to be an integrable infinite-dimensional Hamiltonian dynamical system \cite{Ablowitz2,Ablowitz3,Ablowitz4}. As such, it admits an infinite number of conserved quantities. Of particular interest is the conservation of the so-called ``quasi-power" 
\begin{equation}
\label{Gamma_1_scalar}
\int_{-\infty}^\infty \psi (x,t) \psi^*(-x,t) dx = \Gamma_1\;.
\end{equation}
Furthermore, in \cite{Ablowitz1} a one soliton solution was obtained in the form of a breathing pure one soliton
\begin{equation}
\label{PT-sol}
\psi (x,t) = - \frac{2 (\eta + \bar{\eta})  e^{- 4 i \bar{\eta}^2 t} e^{- 2 \bar{\eta} x} }{1 +  e^{4i (\eta^2 - \bar{\eta}^2)t } e^{-2(\eta + \bar{\eta})x}} \;,
\end{equation}
where $\eta , \bar{\eta}$ are positive constants. Interest in wave propagation in $PT$ symmetric media has been at the forefront of research in physics and mathematics 
\cite{Musslimani2,Musslimani1,Musslimani3,Musslimani4,Musslimani5,yang_review,Cole_Musslimani,Ruter,Konotop,Kevrekidis,Kevrekidis2,Kevrekidis3, Holger1, Holger2, Holger3, yang, Malomed_PT}. 
Here, we show how to use and implement the renormalization scheme to obtain a time-periodic soliton solution. 
Solutions to Eq.~(\ref{PTNLS}) corresponding to initial condition (\ref{IC}) are then computed from the iterative scheme given in Eq.~(\ref{phi-int-iterate-Four-gen}).
The renormalization factor $R(t)$ is computed from the conservation law (\ref{Gamma_1_scalar}) and is given by (assuming it is real)
\begin{equation}
\label{R-nonloc}
R^2(t) = \frac{\Gamma_1 }{\int_{-\infty}^\infty \phi(x,t) \phi^*(-x,t) dx} \; .
\end{equation}
Note that the quantity in the denominator is real valued but not necessarily of a definite sign.

An alternative formula for the renormalization that assumes it be {\it complex} can be derived in a manner similar to what we presented for the classical NLS equation. To do so,
we substitute $\psi (x,t) = R(t) \phi(x,t)$ into the nonlocal NLS equation and, after some algebra, find
\begin{equation}
\label{nonlocal_complex_R}
R(t) = \frac{R(0) \alpha(0)}{\alpha(t)} \exp\left[ - 2 i \Gamma_1  \int_0^t \frac{\gamma(\tau)}
{\zeta (\tau ) \alpha(\tau)} d \tau \right] \; ,
\end{equation}
where $\alpha (\tau)$ is defined in (\ref{alpha}) and
\begin{equation}
\gamma (\tau) = \int_{-\infty}^\infty \phi^2(x,\tau) \phi^*(-x,\tau) dx\; ,
\end{equation}
\begin{equation}
\zeta (\tau) = \int_{-\infty}^\infty \phi(x,\tau) \phi^*(-x,\tau) dx\; .
\end{equation}
The value of $R(0)$ can be obtained by multiplying Eq.~(\ref{IC}) by $f^*(-x)$ and integrating over the entire spatial domain 
\begin{equation}
R(0) =  \frac{\Gamma_1}{\int_{-\infty}^\infty f^*(-x) \phi(x,0) dx} \; .
\end{equation}
We have implemented the time-dependent spectral renormalization scheme to solve the initial boundary value associated with the nonlocal $PT$ symmetric NLS Eq.~(\ref{PTNLS}) corresponding to initial condition given by
\begin{equation}
\label{PT-IC}
f (x) = - \frac{2 (\eta + \bar{\eta}) e^{- 2 \bar{\eta} x} }
{1 +  e^{-2(\eta + \bar{\eta})x}} \;.
\end{equation}
The iteration scheme is seeded with an initial guess 
$\phi_1 (x,t)=f(x).$ After some number of iteration, the method converged and locked on the breathing soliton given in (\ref{PT-sol}). In order to quantify these numerical results we compare our converged solution to the exact one using the global max error given in (\ref{max_norm}). Additionally, since we derive the renormalization factor directly from conservation of quasi-power (\ref{R-nonloc}) we also measure the error in the quasi-power via the quantity 
\begin{equation}
\label{P_nonlocal}
\mathsf{P}_{\rm nonlocal}(t) = \int_{-\infty}^{+\infty} dx \psi_{\rm num}(x,t) \psi_{\rm num}^*(-x,t) - \Gamma_1\; .
\end{equation}
\begin{figure} [h]
\includegraphics[scale=.43]{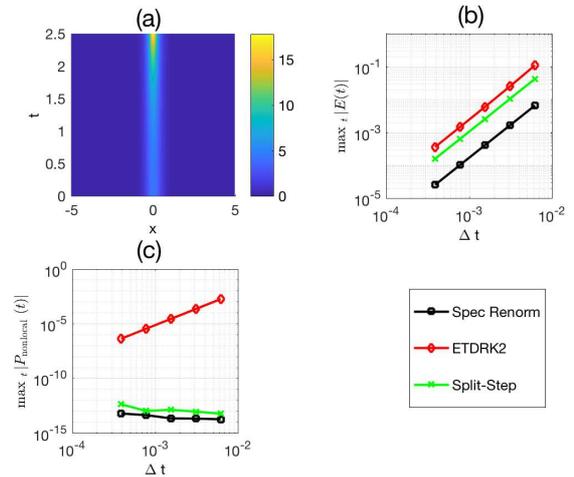}
\caption{(a) A top view in space and time of the solution intensity $|\psi(x,t)|^2$. Global max error of the numerically calculated (b) solution and (c) $\Gamma_1$. Curves correspond to numerical solutions obtained through the spectral renormalization, exponential time-differencing Runge-Kutta (ETDRK2), and split-step methods. Conservation of ``quasi-power" given in Eq.~(\ref{R-nonloc}) is used to determine $R(t).$ The computational parameters are: $ \eta = 1.1, \bar{\eta}= 1, T = 2.5, L =50, N = 1024$.}
\label{FIG6}
\end{figure}
A summary of our findings using the renormalization factor defined in Eq.~(\ref{R-nonloc}) is shown in Fig.~\ref{FIG6}. For this set of parameters, namely $\eta \not = \bar{\eta}$, the soliton mode approaches a singularity at time $t_S = \pi / \left[4 (\eta^2 - \bar{\eta}^2) \right] \approx 3.74$. The solution intensity is shown in Fig.~\ref{FIG6}(a) where the soliton peak is rapidly growing in time. We next compare the numerical solutions obtained from our scheme with other second-order accurate time-integration techniques. Figure~\ref{FIG6}(b) reveals that all the methods exhibit second-order convergence. Particularly, in Fig.~\ref{FIG6}(c) we display the error in the quasi-power (here given by $\Gamma_1 = 4.2$) obtained from three different methods. The split-step and spectral renormalization methods are found to preserve the quasi-power exactly, whereas the error computed using the ETDRK2 method is many orders of magnitude larger. Lastly, we compare the real (\ref{R-nonloc}) and complex (\ref{nonlocal_complex_R}) versions of the renormalization factor. The two different forms of $R(t)$ are shown in Fig.~\ref{FIG7} for identical parameters. As with the classical NLS above, when $R$ is real the magnitude rapidly decays to zero, while when $R$ is complex it is observed to oscillate in time.
\begin{figure} [h]
\includegraphics[scale=.4]{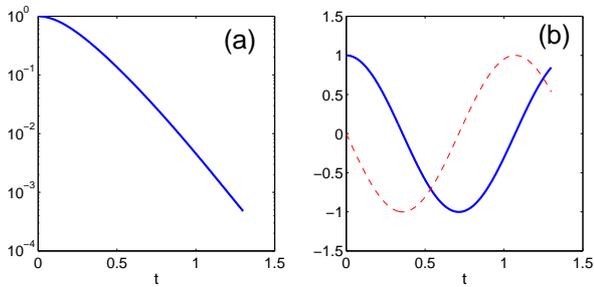}
\caption{Dependence of the renormalization factor on time. (a) Real and (b) complex versions of the renormalization factor upon convergence of the spectral renormalization method. The real (solid blue line) and imaginary (dashed red line) parts are shown. The parameters are the same as those in Fig.~\ref{FIG6} with $T = 1.3$ and $M = 400$ time points.}
\label{FIG7}
\end{figure}
\section{Dissipative Case: Burger's equation}
\label{dissipate_sec}
So far we have explained and implemented the time-dependent spectral renormalization algorithm in conservative systems and demonstrated its flexibility in terms of incorporating conserved physical properties into the simulation. In this section, we turn our attention to dissipative evolution equations and apply the scheme on the viscous Burgers' equation 
\begin{equation}
\label{burgers}
\psi_t = - \psi \psi_x + \nu \psi_{xx} \;.
\end{equation}
Here, $\psi = \psi (x,t)$ is a real valued function and $\nu >0$ is the viscosity coefficient. The problem will be considered on the bounded domain $[0,2 \pi]$ with periodic boundary conditions:  $\psi (x+2\pi,t)=\psi (x,t)$ and $\psi_x (x+2\pi,t)=\psi_x (x,t).$ The integration time interval is $[0, T].$ The limit as $\nu \rightarrow 0$ is referred to as the inviscid Burgers' equation. It was first proposed by Burgers \cite{Burgers} as a model for turbulent flows and later shown to play a fundamental role in many branches of the nonlinear sciences. This equation exhibits rich mathematical structures such as finite time blow-up (singularity), shock formation and wave breaking \cite{Ablowitz_book}. Physically speaking, it finds wide applications in gas dynamics, acoustics and traffic flows to name a few  \cite{Whitham}. The viscous Burgers' ($\nu\ne 0$) is often used in fluid mechanics as a simplified version of the Navier-Stokes equation in one space dimension under the assumption of incompressibility and no pressure gradient. Importantly, Eq.~(\ref{burgers}) can be solved exactly with the help of the so-called Cole-Hopf transformation \cite{Cole, Hopf}. Indeed, if $\psi (x,t)$ solves Eq.~(\ref{burgers}) then the function $u(x,t)$ defined by 
\begin{equation}
\label{burgers_soln1}
\psi (x,t) = - 2 \nu \frac{u_x(x,t)}{u(x,t)} \; ,
\end{equation}
satisfies the linear heat equation 
\begin{equation}
\label{heat_eq}
u_t(x,t) = \nu u_{xx}(x,t)\;.
\end{equation}
A spatially periodic exact solution to Eq.~(\ref{burgers}) corresponding to the initial condition
\begin{equation}
\label{IC-burgers}
\psi (x,0) = f(x) = -\frac{\nu \cos (x)}{1 + \frac{1}{2} \sin\left( x\right)} \; ,
\end{equation}
is given by
\begin{equation}
\label{burgers_soln2}
\psi_{\rm ex} (x,t) = - \frac{\nu \cos (x) e^{- \nu t}}{1 + \frac{1}{2} \sin\left( x\right)e^{- \nu t}} \; .
\end{equation}
The viscous Burgers' equation is dissipative in nature. Indeed, multiplying Eq.~(\ref{burgers}) 
by $\psi$ and integrating over the entire spatial domain we find 
\begin{equation}
\label{dissipate_rate}
\frac{d }{d t} \int_{0}^{2 \pi} \psi^2 (x,t) dx = - 2 \nu \int_0^{2 \pi} \psi^2_x(x,t) dx \; .
\end{equation}
This relation describes the rate at which the total energy of the system is dissipated. 

With all this at hand, the initial boundary value problem (\ref{burgers}) corresponding to initial condition (\ref{IC-burgers}) is solved using the iterative scheme given in Eq.~(\ref{phi-int-iterate-Four-gen}).
For the viscous Burgers' equation we have the linear operator $\hat{\mathcal{L}}(k) = -\nu k^2$ and nonlinearity $\mathcal{N} = - \left( \psi^2 \right)_x / 2 $.
To derive a formula for the renormalization factor, we substitute $\psi (x,t) = R(t) \phi (x,t)$  
into (\ref{dissipate_rate}); integrate over the entire spatial domain and obtain
\begin{equation}
\label{renorm_dissipate_rate}
\frac{d (\theta_1 R^2) }{dt} = - 2 \nu \theta_2 R^2   \; ,
\end{equation}
where
\begin{equation}
\label{beta}
\theta_1(t) = \int_{0}^{2 \pi} \phi^2(x,t) dx \not = 0 \; , 
\end{equation}
\begin{equation}
\label{gamma}
\theta_2(t) = \int_{0}^{2 \pi} \phi^2_x(x,t) dx \; .
\end{equation}
Solving Eq.~(\ref{renorm_dissipate_rate}) for nontrivial solutions we obtain
\begin{equation}
\label{lambda_burgers_define}
R^{2}(t)=  \frac{ R^2(0) \theta_1(0) }{\theta_1(t)} 
\exp \left( - 2 \nu \int_0^t \frac{\theta_2(\tau)}{\theta_1(\tau)}~d\tau \right)  \; .
\end{equation}
Unlike the NLS case, the quantities $\theta_j(t), j=1,2$ are non-negative and real.
To evaluate the cumulative time integral in Eq.~(\ref{lambda_burgers_define}) a second-order accurate trapezoidal method is used.
\begin{figure} [h]
\includegraphics[scale=.43]{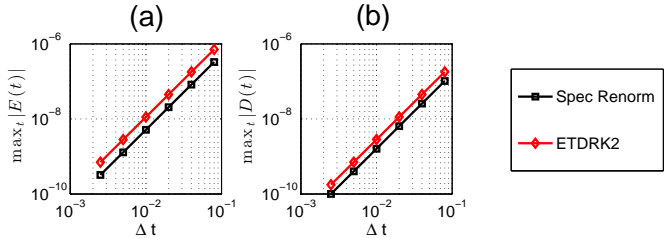}
\caption{(a) Global max error of the numerically calculated solution. (b) Global (time-dependent) dissipation rate error [defined Eq.~(\ref{D_burgers})] in comparison to the exact rate.
Curves correspond to numerical solutions obtained through the spectral renormalization and exponential time-differencing Runge-Kutta (ETDRK2) methods. The computational parameters are: $\nu = 1/10, T = 10, L =2 \pi, N = 64$.}
\label{FIG8}
\end{figure}
Since we are directly enforcing dissipation of energy in Eq.~(\ref{dissipate_rate}) we are interested to learn how the energy in the numerical solution compares to that of the exact one. For that purpose, we define the quantity
\begin{equation}
\label{D_burgers}
\mathsf{D}(t) = \int_{0}^{2 \pi} dx \left( \psi^2_{\rm num}(x,t)  - \psi^2_{\rm ex}(x,t) \right) \; ,
\end{equation}
where $\psi_{\rm num}(x,t)$ is the numerical solution obtained from iterating 
Eq.~(\ref{phi-int-iterate-Four-gen}) using the initial condition (\ref{IC-burgers}). The error convergence rates are shown in Fig.~\ref{FIG8}. Again, the global solution error in Fig.~\ref{FIG8}(a) is observed to converge at a second-order rate. This is also the case for the
ETDRK2 scheme. Next, the error in the (time-dependent) dissipation rate is compared with the exact. Doing so reveals that, unlike the time-independent conserved quantities above, the dissipation rate is not numerically exact, but instead converges at second-order rate [see Fig.~\ref{FIG8}(b)]. The reason for this is that the time integral in Eq.~(\ref{lambda_burgers_define}) is approximated by a second-order accurate quadrature method. A similar convergence rate is found for the Runge-Kutta scheme.
\section{Conclusions}
\label{conclude}
In this paper, we have presented a new numerical method to simulate evolution equations which we term here the time-dependent spectral renormalization scheme.  
The proposed scheme is rooted in the ({\it time-independent}) spectral renormalization method introduced in 2005 by Ablowitz and Musslimani \cite{spec_renorm_AM} to numerically compute stationary nonlinear bound states for certain nonlinear boundary value problems. In this regard, the proposed method can be thought of as an extension to the time domain. The idea here is to rewrite the given evolution equation (formulated in an ordinary or partial differential equation form) into an integral equation. The solution is then viewed as a fixed point in both space and time, rather than being a solution to an evolution equation. The resulting integral equation is then numerically solved using a simple renormalized fixed-point iteration. Convergence of the method is achieved by introducing a {\it time-dependent} renormalization factor which is numerically computed from the {\it physical properties} of the governing evolution equations. This novel time-dependent spectral renormalization scheme has the ability to incorporate physics into the numerical simulations and allows the numerical time integration to ``keep in touch" with the original evolution equation. The proposed method is applied to two benchmark problems: the nonlinear Schr\"odinger and the Burgers' equations each of which being a prototypical example of a conservative and dissipative system respectively.
\section{Acknowledgment}
The work of J.T.C. and Z.H.M was supported in part by NSF grant
number DMS-0908599. Z.H.M thanks the Lady-Davis Trust for the financial support during his visit to the Technion-Israel Institute of Technology.

\end{document}